\documentstyle[prl,aps,twocolumn]{revtex}
\begin{document}
\title{Magneto-optical properties of charged excitons in quantum dots}
\author{A. O. Govorov$^{1,2,3}$, C. Schulhauser$^{1}$, D. Haft$^{1}$, A. V. Kalameitsev$^{3}$, A. Chaplik$^{3}$, R. J. Warburton$^{4}$,\\ K. Karrai$^{1}$, W. Schoenfeld$^{5}$, J. M. Garcia$^{6}$, and P. M. Petroff$^{5}$}
\address{
$^{1}$Center for NanoScience and Sektion Physik, Ludwig-Maximilians-Universit\"{a}t,\\
Geschwister-Scholl-Platz 1, 80539 M\"{u}nchen, Germany\\
$^{2}$Department of Physics and Astronomy and CMSS Program, Ohio University, Athens, OH 45701\\
$^{3}$Institute of Semiconductor Physics, RAS, Siberian Branch, 630090 Novosibirsk, Russia\\
$^{4}$Department of Physics, Heriot-Watt University, Edinburgh EH14 4AS, UK\\
$^{5}$Materials Department, University of California, Santa Barbara, California 93106\\
$^{6}$Instituto de Microelectronica de Madrid, CNM-CSIC Isaac Newton, 8, PTM, 28760 Madrid, Spain}

\date{\today}

\maketitle

\begin{abstract}
We present both experimental and theoretical results on the influence of a magnetic field on excitons in semiconductor quantum dots. We find a pronounced difference between the strong and weak confinement regimes. For weak confinement, the excitonic diamagnetic shift is strongly dependent on surplus charge, corresponding to a reversal in sign of the conventional diamagnetic shift for neutral excitons. In this limit, we argue that the optical properties of excitons with two or more extra electrons are fundamentally different to those of the neutral exciton and trion.
\end{abstract}

\narrowtext

A semiconductor quantum dot (QD) represents an ideal model system for the investigation of quantum mechanical electron-electron interactions. This is because Coulomb blockade allows electrons to be added or removed one by one simply with a gate electrode \cite{Drexler,Tarucha}. As a result, the electrical \cite{Tarucha}, optical \cite{Warburton1}-\cite{Findeis} and magnetic properties \cite{Wiel} are tunable. An exciton complex consists of a hole bound to the electrons in a QD. The spatial extent of the excitonic wave function reflects the joint effects of the QDs confinement potential and the Coulomb interactions and can be probed by applying a magnetic field, $B$. For neutral excitons, the exciton energy increases quadratically with $B$, the so-called diamagnetic shift, with a curvature proportional to the area of the wave function \cite{Seisyan}. However, the behavior of charged excitons is less well-known and potentially much more interesting because of the more elaborate Coulomb interactions.

Here, we address both experimentally and theoretically the effect of electron charging on the excitonic diamagnetic shift. We show that each additional charge leads to a new {\em paramagnetic} contribution. Unlike paramagnetism in solids and atoms, we propose that the QD paramagnetism is a signature of strong Coulomb interactions. The multiply-charged exciton complexes we investigate are not stable in either homogeneous bulk semiconductors or quantum wells and so by turning to QDs we have entered a new regime where the Coulomb interactions can dominate the response to a magnetic field. To analyse this case, we introduce the concept of Wigner molecules for charged excitons.

For the experiments, we used self-assembled InAs QDs where it is well known that excitons recombine efficiently through photon emission \cite{Marzin} and that their charge can be controlled electrostatically \cite{Drexler}. The InAs QDs are grown by molecular beam epitaxy and are embedded 25 nm above a highly doped GaAs layer, the back contact, and 150 nm below a Schottky gate on the surface. The QDs are grown in the Stranski-Krastanow mode, giving at our growth temperature 520 $^{\rm o}$C lens-shaped QDs. We then deposit 1 nm of GaAs before annealing the sample at the growth temperature for 1 min \cite{Garcia}. Ensemble photoluminescence (PL) experiments show that the annealing step produces a multi modal distribution of dots: there are two dominant PL bands, one centred at 1.31 eV (the red-band) and one at 1.34 eV (the blue band). We exploit this property here as it allows us to study the effects of different confinement strengths within the same sample. In order to excite the PL, we generate carriers in the wetting layer with a 822 nm laser diode. We measure the PL from single quantum dots by processing 300 nm diameter apertures in the otherwise opaque metal gate, collecting the PL with a confocal microscope at 4.2 K and up to $B=9$ T. In all the experiments, the pump intensity was low enough that emission from biexcitons was undetectable.

The basic experiment is to measure the PL energy as a function of gate voltage, $V_{g}$, and magnetic field. We observe a red-shift in the PL whenever a single electron tunnels from the back contact into the QD, and from the Coulomb blockade plateau in the PL we can determine the charge of the exciton \cite{Warburton1}. The QDs emitting near 1.26 eV on the red side of the ensemble PL acquire one extra electron at $V_{g} \sim -0.65$ V and can be filled with as many as 3 electrons before the charge spills out into the wetting layer. For the QDs emitting near 1.37 eV on the blue side of the ensemble PL, the confinement potential is substantially weaker such that the charging threshold moves to $V_{g}\sim -0.15$ V, and only one extra electron can be added.

A typical $B$-dependence of the PL is shown in Fig.\ 1. The PL line splits into two in magnetic field through the Zeeman effect. The splitting is 120 $\mu$eV/T, varying by $\pm 30$ $\mu$eV from dot to dot, without any measurable dependence on excitonic charge. In order to analyse the overall up-shift of the PL in $B$, we plot the upper (lower) branch against positive (negative) $B$-values. Such a representation reveals the quadratic dependence of the PL peak shift: $\Delta E_{PL}=g_{ex}\mu_{B}B/2+\alpha B^{2}$ where $g_{ex}$ is the Land\'{e} factor and $\mu_{B}$ the Bohr magneton. We focus on $\alpha$ which we have measured for about 20 different QDs.

Fig.\ 2 shows the PL dispersion of two different QDs, one from the red-band of the ensemble PL, and one from the blue-band, for different excitonic charges. The red-band QD has $\alpha =10 \pm 1$ $\mu$eV/T$^{2}$, independent of the excitonic charge, and we find that this is the case for all the investigated dots in the red-band. In very clear contrast however, the dots in the blue-band have the remarkable property that the diamagnetic shift reduces with the addition of one electron. An example is shown in Fig.\ 2 where the neutral exciton has $\alpha= 16.6$ $\mu$eV/T$^{2}$, the singly charged exciton 8.7 $\mu$eV/T$^{2}$. In other words, the extra electron makes a paramagnetic contribution of $\alpha=-7.9$ $\mu$eV/T$^{2}$ to the overall diamagnetism. We argue in the following that the paramagnetism is a consequence of Coulomb interactions in the QD.

In order to understand these experimental results, we present generic calculations in two different limits, strong and weak confinement. The aim is to achieve a qualitative understanding of the experimental results. A complete quantitative agreement is probably only possible using the exact confinement potentials which are generally unknown for self-assembled QDs and it is not our purpose to explore this issue here. Importantly, we reach some wide-ranging conclusions which are independent of the form of the potential. For simplicity, we take a two-dimensional (2d) parabolic potential for both electrons (e) and holes (h) of the type $V_{e(h)}=m_{e(h)}\Omega_{e(h)}^{2}r^{2}/2$ where $\Omega_{e(h)}$ are the single particle frequencies, $m_{e(h)}$ the effective masses, and $r$ is the spatial in-plane coordinate \cite{Jacak}. In the strong confinement regime, the single particle energies dominate over the Coulomb energies, such that the Coulomb energies can be treated as a perturbation \cite{Warburton2}. The diamagnetic shift of the $n$-times negatively charged exciton X$^{n-}$ is proportional to $B^{2}$ in the limit where the electron (hole) cyclotron frequencies $\omega_{e(h)}^{cr}=eB/m_{e(h)} \ll \Omega_{e(h)}$. The results for $\alpha$ are plotted in Fig.\ 3 showing how $\alpha$ depends on excess charge, with the changes becoming more important as the confinement weakens. However, typical confinement energies for strongly confined InAs quantum dots are $\hbar\Omega_{e}\sim 30$ to 50 meV for which the $\alpha$'s for the X$^{0}$, X$^{1-}$ and X$^{2-}$ excitons differ by only $\sim 10$ \%. The magnitude of this effect is comparable to our experimental resolution in $\alpha$. The diamagnetic shift of the red-band QDs are therefore consistent with the predictions of theory in the strong confinement limit.

In the other regime, weak confinement, the Coulomb energies dominate over the single particle energies. The magnetic shifts of a freely moving X$^{0}$ and X$^{1-}$ were treated in ref.s \cite{Govorov1,Stebe}. It was found that the X$^{1-}$ has a negative magnetic dispersion, arising from the cyclotron motion: the electron mass in the final state is much less than the trion mass in the initial state. Experimentally, a weak paramagnetic dispersion for the X$^{1-}$ in a quantum well has been observed in fields of about 1 T \cite{Shields,Finkelstein}. We now consider the case of an X$^{1-}$ in a parabolic confinement potential. The effective potential is $2 V_{e}+V_{h}$ and so the center of mass motion of a trion is described by a harmonic wave function with radial and angular quantum numbers, both of which are 0 in the ground state. Using the single particle Fock-Darwin spectrum, we determine the X$^{1-}$ exciton energy to be $E^{1-}=E_{tr}+\hbar\left[ \Omega_{tr}-\Omega_{e}(2N+1)\right]$, where $E_{tr}$ is the free 2d trion energy, $N$ is the radial quantum number of the electron left in the final state, and $\Omega_{tr}^{2}=(2 m_{e}\Omega_{e}^{2}+m_{h}\Omega_{h}^{2})/(2m_{e}+m_{h})$. The interband selection rule on the envelope function dictates that the final state after photon emission should also have a zero angular momentum. Based on this, we find a $B^{2}$-dispersion of the PL emission energies with $\alpha_{N}^{1-}=(\hbar e^{2}/8)\left[ 1/m_{tr}^{2}\Omega_{tr}-(1+2N)/m_{e}^{2}\Omega_{e}\right]$, which is negative for all $N$ because of the inequality $m_{tr}\gg m_{e}$. The physical reason for $\alpha_{N}^{1-}<0$ is that the final state is more extended than the initial state. The magnetic field dispersion for $N=0$ is shown in the left panel to Fig.\ 4. The intensity of the PL emission lines depends on an overlap integral of the trion and electron wave functions. With increasing radial quantum number $N$, the overlap and hence PL intensity decrease rapidly (right panel, Fig.\ 4). The lines with $N>0$ originate from a shake-up process in which the final state is an excited electron state. Such processes have already been investigated in the tunneling \cite{Ashoori} and PL spectroscopy of 2d systems \cite{Jacak,Finkelstein,Goldberg,Wojs}.

Experimentally, the diamagnetic curvature of X$^{1-}$ in a blue-band QD is much smaller than that of X$^{0}$, but the dispersion is not paramagnetic. Also, we do not observe the shake-up peaks. Our explanation is that the blue-band QDs are in an intermediate regime. Interpolating between the two limits of our theory, in the intermediate regime the shake-up peaks will be weak and the diamagnetic contribution small and positive, giving us qualitative agreement with the experiment.

In the weak confinement regime, there are pronounced changes on going from the X$^{0}$ to the X$^{1-}$: the appearance of paramagnetism and shake-up peaks. We find further changes on going from X$^{1-}$ to X$^{2-}$. The X$^{2-}$ can be viewed as an electron plus a trion, namely a Wigner molecule \cite{Govorov2}. In order to determine the classical coordinates of the trion and the electron, we minimized the classical energy of the Wigner molecule. The result is that the classical distance $d_{e}^{i}$ of the electron to the dot center is a factor $\beta = m_{tr}\Omega_{tr}^{2}/m_{e}\Omega_{e}^{2}$ larger than $d_{tr}^{i}$, the corresponding distance for the center of mass of the trion. This is because the trion confinement potential $V_{tr}$ is stronger than the electron confinement potential $V_{e}$. In a similar way, we determined the classical configuration of the two electrons left in the final state after photon emission. Their classical coordinate $d_{e}^{f}$ in the ground state obeys $d_{e}^{i}>d_{e}^{f}>d_{tr}^{i}$ as seen in Fig.\ 4. The wave functions of the initial and final states are peaked at the classical coordinates and decay exponentially along the radial axis away from their peak positions. Since $\beta \ne 1$, the classical coordinates for the initial and final states are different. This gives a reduction in the PL intensity, but there is still a significant oscillator strength because the wave functions of the final excited states extend far enough. An example is shown in Fig.\ 4. This is the origin of the shake-up process in the PL of QDs. To determine the magnetic dispersion of the PL, we modeled the initial ground state as a Wigner molecule and computed numerically all the final states of two electrons. The total energy of the initial state includes the classical energy, the rotation energy and the quantization in the self-consistent local potential minima, but neglects exchange effects within the Wigner molecule. $\alpha^{2-}$ is estimated from the distance $\Delta d \sim | d_{tr}^{i}-d_{e}^{f}| \sim |d_{e}^{i}-d_{e}^{f}|$. The final state of the most intense PL line has a radial extent comparable to $\Delta d$ since it corresponds to the maximum overlap with the initial state. From the diamagnetic shift of the final state which is $\sim (e^{2}B^{2}/8m_{e})\Delta d^{2}$, we determine an approximate paramagnetic shift $-(e^{2}B^{2}l_{e}^{2}/8m_{e})(l_{e}/a_{o}^{*})^{2/3}$ for the most intense PL line. Here $l_{e}=(\hbar /m_{e}\Omega_{e})^{1/2}$, the electron zero-point motion, and $a_{o}^{*}$ is the Bohr radius. Assuming $m_{e}\ll m_{tr}$ we see that the curvature of this X$^{2-}$ dispersion is about a factor $(l_{e}/a_{o}^{*})^{2/3}$ larger than that of X$^{1-}$ since in the weak confinement regime $l_{e} \gg a_{o}^{*}$. This is shown in Fig.\ 4. Note that the jumps in the PL energy for X$^{2-}$ originate from changes in the angular momentum in the ground state, as in a two-electron quantum dot \cite{Jacak}. The prediction for X$^{2-}$ in the weak confinement limit is that shake-up processes dominate the PL, and that there is a large, Coulomb-induced paramagnetic dispersion. Similar arguments apply also to more highly charged excitons, as shown for X$^{3-}$ in Fig.\ 4 (middle panel).

In conclusion, we report measurements of the diamagnetic shift of charged excitons in quantum dots spanning the strong to intermediate confinement regimes. In the strong confinement regime, the diamagnetic shift is small, and independent of charge. In the intermediate regime, the diamagnetic shift of a neutral exciton is larger, but there is a significant decrease with the addition of a single electron. Theoretical models are developed to understand these results, and we extend the theory to the case of weakly confined doubly charged excitons where we predict pronounced Coulomb interactions, leading to an unusual paramagnetic behavior and to strong shake-up processes in the emission. While these predictions pertain to low fields for InAs QDs ($B \le 2$ T), they apply to higher fields for II-VI materials where the trion is more strongly bound \cite{Astakhov}. The novelty of these results is based on the fact that highly charged excitons do not exist in bulk semiconductors and quantum wells.

We acknowledge S.\ Ulloa for fruitful discussions. This work was supported by the DFG under SFB348 and EPSRC. One of us, A.O.G., acknowledges financial support from the Rufus Putnam Visiting Professorship and the Volkswagen Foundation.

\newpage
\begin{figure}
\caption{Left: gray scale plot of the photoluminescence (PL)
intensity against magnetic field, $B$. Black corresponds to 240
counts in 120 seconds on the detector; white is the background
signal. Right: the peak positions of the upper and lower branches
shown left plotted against positive and negative $B$,
respectively. The solid line is a fit of the energy to a second
order polynomial in $B$.}
\end{figure}

\begin{figure}
\caption{Left: diamagnetic shift against magnetic field for the PL of a quantum dot emitting in the 1.31 eV band. The three symbols correspond to X$^{0}$, X$^{1-}$ and X$^{2-}$ excitons. Right: diamagnetic shift for a quantum dot emitting in the 1.34 eV band. The paramagnetic contribution due to charging is demonstrated by plotting the energy of X$^{1-}-$X$^{0}$.}
\end{figure}

\begin{figure}
\caption{Diamagnetic curvature $\alpha$ plotted against the electron quantization energy $\hbar \Omega_{e}$ for different excitonic charges. The curves are calculated in the strong confinement limit within first order perturbation theory. The dashed line is the diamagnetic shift of a non-interacting electron-hole pair.}
\end{figure}

\begin{figure}
\caption{Left: the energy dispersion of the most intense X$^{0}$, X$^{1-}$ and X$^{2-}$ PL lines calculated in the weak confinement limit. Middle: classical positions of the trion and excess electrons in the dot before and after recombination for X$^{1-}$, X$^{2-}$ and X$^{3-}$. The jumps in energy have been reduced by a factor of 10. Right: PL spectra showing the shake-up peaks. $\omega_{\rm main}$ is the frequency of the strongest emission, $R_{y}^{*}$ is the effective Rydberg, $\hbar \Omega_{e}$ is the electron quantization energy, and $\hbar \Omega_{e}/R_{y}^{*}\ll 1$ for weak confinement.}
\end{figure}

\end{document}